# Smart Geographic object: Toward a new understanding of GIS Technology in Ubiquitous Computing

Sakyoud Zakaria [1,2], Gaëtan Rey [2], Eladnani Mohamed [1]
Stéphane Lavirotte [2], El Fazziki Abdelaziz[1], Jean-Yves Tigli [2]

[1]Cadi Ayyad University of Marrakech - Faculty of science - Morocco.

[2]University Nice Sophia Antipolis - I3S Laboratory - France.

**Abstract**
One of the fundamental aspects of ubiquitous computing is the instrumentation of the real world by smart devices. This instrumentation constitutes an opportunity to rethink the interactions between human beings and their environment on the one hand, and between the components of this environment on the other. In this paper we discuss what this understanding of ubiquitous computing can bring to geographic science and particularly to GIS technology. Our main idea is the instrumentation of the geographic environment through the instrumentation of geographic objects composing it. And then investigate how this instrumentation can meet the current limitations of GIS technology, and offers a new stage of rapprochement between the earth and its abstraction.
As result, the current research work proposes a new concept we named Smart Geographic Object SGO. The latter is a convergence point between the smart objects and geographic objects, two concepts appertaining respectively to the Internet of Things, and geographical modeling of the Earth.

**Keywords:** *GIS, Ubiquitous Computing, Geographic object, Smart object.*

## 1. Introduction

The geographic environment is too complex to be understood in its entirety. So, when dealing with a problem in this environment, we should focus on specific aspects of our concerns. For example, a geographical interrogation may concern a geographical area, geographic objects in this area, or just some specific properties of these objects. GIS technology has always provided an operating framework to store and interrogate geographic data. It has enabled users to have partial views of the geographic environment and its objects [31]. Hence, objects and their relevant characteristics can be distinguished in every interrogation.

However, GIS technology faces a number of limitations. First, the acquisition process of geographic data is extremely expensive. Also, the geographic data update is more or less frequent depending on the areas accessibility.

GIS, on the other hand, inherits the location technologies limitations, such as the imprecision of localization and the signal availability. In fact, this limitation translates a generic and classic challenge in geographic science, which is the gap between the geographical environment and its abstraction. In other words, the absence of a perfect and exact representation of the Earth can cause the imprecision of the geographic interrogations results.

Science in general and computer science in particular, has reduced the gap between geographical environment and its representation. This rapprochement constitutes the history of the geographical science, especially the history of the cartography. Our research fits into this evolution process, and proposes a new level of rapprochement exploiting ubiquitous computing's corps concepts.

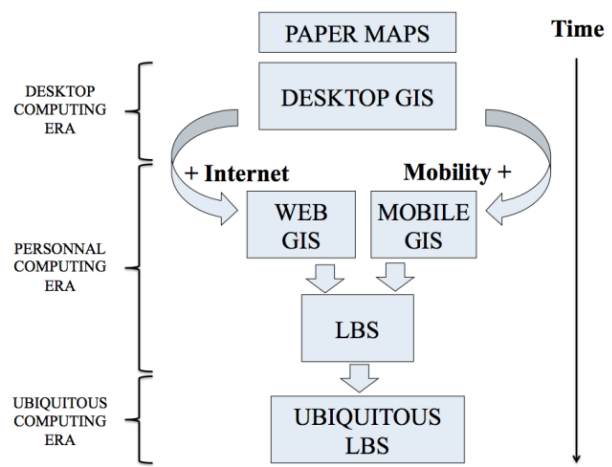

Fig. 1 Evolution of GIS technology in the three eras of computing distinguished by Mark Weiser

To designate the position of our work, we illustrate in the figure 1 the alignment of GIS evolution according to the three computing eras defined by Mark Wesier in his famous paper "The computer for the 21st century" [1].

Actually, we can distinguish three generations of GIS in the computing age:
- *Desktop GIS* that were intended for professional use in specialized organizations and labs. They still exist, but they are often playing the role of geographic servers.
- *Mobile and Web GIS* who have democratized the use of GIS technology benefiting from the improvement of mobile computing and Internet; and who gave birth to location-based services LBS. They open a new area to provide value-added services based on location; like advising traffic conditions, routing information …[17]
- *Ubiquitous LBSs* that will characterize the era of ubiquitous computing, where GIS will operate in symbiosis with their environment. Our current research work belongs to this third category.

Unlike many existing works, we will not be limited in our study to mobility and context-awareness, but we will focus on making the geographic environment a part of the GIS applications functioning. Our proposition in other words aims to provide the geographic environment with exchange capacities with its physical and virtual environments.
The geographic environment is a set of geographic object. To make the geographic environment smart, we should make the geographic objects composing it enough smart, to communicate with each other one side and with their physical and virtual environment on the other.

Our current work involves hence a binary understanding including smart objects and geographic objects, two concepts appertaining respectively to the Ubiquitous computing, and geographical modeling of the Earth. We gathered these tow concepts in one concept we called Smart Geographic Object SGO, and it aims to provide geographic objects with exchange capacities with their physical and virtual environments. The Smart Geographic Objects purpose is also to raise new forms of interaction with the geographical environment; and give responses to some crucial GIS technology limitations.

This paper is organized as follows:
- *Section 2: presents the Concept of Smart Geographic object, and the reasoning behind this concept.*
- *Sections 3: presents a scenario illuminating some of our work motivations.*
- *Section 4: presents a generic conceptual architecture for Smart Geographic Objects.*
- *Section 5: Will demonstrate how SGO solves the GIS limitations identified through the motivation scenario*
- *Section 6: Will situate our work compared with some related works before concluding.*

## 2. Smart Geographic Object

One of the major questions that enthused this work is: what are the specific features of GIS applications in ubiquitous computing? Most work in our state of the art answers to this question by giving tow features, the first is mobility and the second is context-awareness. Actually, the previous features are not specific to GIS applications, and every application in ubiquitous computing may have the same aspects. This is why we thought to situate GIS applications in a global ubiquitous computing architecture to have a conscientious answer to the question.

A global architecture of ubiquitous computing, such as that proposed by Stefan Poslad [30], defines three basic architectural design patterns for ubiquitous computing: smart devices, smart environment and smart interaction. The notion of "smart" here means that the components are active, networked and can operate to some extent autonomously:
- *Smart devices* are for example personal computers, mobile phones operating as a single portal to access applications and services. These services may reside locally on the device or remotely on servers.
- *Smart environment* consists of a set of networked devices embedded into the physical world. Unlike smart devices, the devices that compose a smart environment usually execute a single predefined task, for example motion or body heat sensors coupled to a door release and lock control.
- Finally, in order for smart devices and smart environments to support the core properties of ubiquitous computing, *Smart Interactions* are needed to promote a unified and continuous interaction model between ubiquitous applications and the ubiquitous infrastructure.

Figure 2 illustrates the positioning of GIS technology in the previous architecture. Through this positioning, we can clearly remark that the specific features of GIS applications in ubiquitous computing lies in the interaction with an intelligent geographic environment.

It is in fact an interaction with the geographic objects of this environment. Hence, in addition to mobility (smart device) and context-awareness (smart interactions), the reflection on GIS technology in ubiquitous computing must also take in account the role of the potentially smart geographical environment. To shape this reflection we proposed the concept of Smart Geographic Objects, which constitutes -as we see it- the basic building block of a smart geographic environment.
Before formalizing our perception of the Smart Geographic Object, we will present the two concepts composing it,

which are Smart Object and Geographic Object. In fact, we can find an important number of definitions for smart object [3] [4] [5] and geographic object [6] [7] [8] in the literature. So we will present a reading of the two concepts that meet the work.

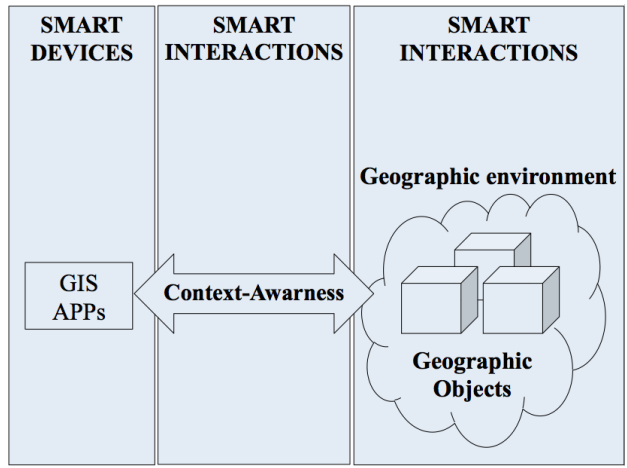

Fig. 2 the positioning of GIS technology in Smart EDI architecture

2.1 Geographic Object

A geographic object is an object in space characterized by well-defined boundaries and properties such as the name and type [10]. A geographic object is not only located in space, but it is intrinsically linked to this space, because it inevitably inherits its structure and its geometric, topological, and climatic properties [11].

A geographic object is abstracted via a set of spatial data. This spatial data include geometrical information and thematic information [10]. Geometric data describe the position and shape of the geographical entity within a reference coordinate system. Thematic data describe the attributes of a geographic entity, excluding its shape and position.

It is difficult to find a standard definition of the geographic object due to the diversity and complexity of geographical entities that exist in the environment. For example, the concept of geographic object according to a geology research group is not the same for a land conservation service. So for each field, a definition of the geographic object should be made to comprise the specificities related to this field. Each definition should also take into account a specific reference system, because the definition of the spatial object may vary by the variation of the level of detail even in the same disciplinary field. The level of detail here is similar to the notion of scale in a paper map.

In our research, we focus on geographic objects in urban area. From a geometrical angle, the geographic objects in urban area can be classified into simple objects and complex objects. There are four simple objects that are actually primitive components of complex objects, the point (E.g valve), the line (E.g a road), the polygon (E.g a border area) and the solid (E.g building). These simple objects can be combined to form complex components (E.g city) [14]. From a thematic angle, the classification of geographic object in urban areas includes the definition of thematic classes (buildings, ponds, urban supply, etc.) and their relationships [15]. This classification is originally made from types recognized by individuals or determined by the authorities.

In addition to geometric and thematic data, we distinguish a third type of data that we call related services. Indeed, a geographic object in urban areas can be a location of a commercial, administrative, or signage activity. A set of services is hence related to this geographic object. Unlike geometric and thematic data that inform static information about the object, the information provided by related services inform about the activity and the behaviour of the geographic object. For example, the related services to a hotel are room availability, rates, promotions, entertainment activities, etc.

Therefore in urban areas, "*a geographic object is a simple or a complex spatial entity characterized by a set of geometric and thematic data plus a set of related services. Geometric data provide the position, the shape and scope. Thematic data reflect the type, classification and relationship with other spatial entities. The related services in turn provide information about the activity of the geographic object*".

2.2 Smart Object

In the new era of ubiquitous computing objects that surround us will be one way or another connected. They will be connected to a large and global network thanks to the development of wireless connection technologies. This network is none other than the Internet of things. And the term Smart Object makes reference to an object appertaining to this network.
A smart object can be defined as a real object increased by technological capabilities. These capabilities allow it to communicate and interact with the environment. As purpose, the concept of smart object aim to provide a real-world object with new functions in addition to its native ones. Hence real-world objects are intended to retrieve, store, transfer and process data without discontinuity between the physical and the digital world.

Technically the implementation of a smart object involves software / hardware instrumentation. The hardware part is the boarding of devices and sensors. The software part is the deployment of software layers to interpret the collected data [28]. To summarize the requirements to increase real world object functions, Karimi and Atkinson [9] propose the following elements:

- Sensing and data collection capability (sensing nodes)
- Layers of local embedded processing capability (local embedded processing nodes)
- Wired and/or wireless communication capability (connectivity nodes)
- Software to automate tasks and enable new classes of services
- Remote network/cloud-based embedded processing capability (remote embedded processing nodes)

In our work, the real-world objects we are trying to make "smart" are geographic objects like buildings and roads. It's about equipping geographic objects with sensing and connectivity nodes, plus local and remote processing capabilities. The idea here is making "a smart object at a big scale", taking into account the specificity of geographic world.

The next section defines the smart geographic object Based on the presentation of the geographic object and smart object concepts.

2.2 Smart Geographic Object

The Smart Geographic Object is born from an intersection between smart objects and geographic object concepts (Figure 3). This fusion's merger is to apply fully ubiquitous computing concepts to the geographic domain.

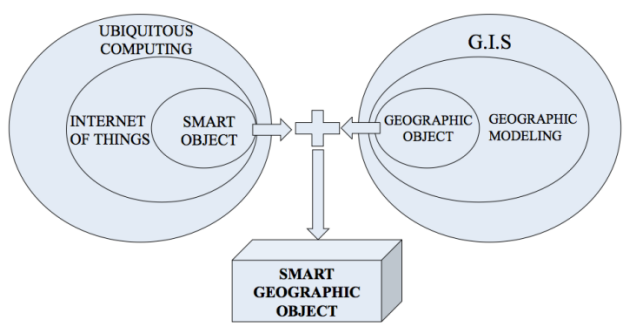

Fig 3 The SGO an intersection between geographic object and smart object concepts

First, we defined the geographic object in urban areas as a spatial entity characterized by geometric and thematic data, in addition to related services. On the other hand, we specified that a smart object is a real-world object augmented with technological capabilities that extends its native functions. Across the foregoing elements the smart geographic object can be defined as:

"*Geographic entities in urban area augmented with capture, processing and connection capabilities, able to communicate its description (geographic data, thematic data, and related services) to the environment. And receives in the other direction information from the environment to update this description*"

## 3. Motivation scenario

To explain more clearly the motivation and the interest of our proposition, we illustrate then formalize some of GIS limitations that still challenging geographic science. The scenario describes the tourism tour of Dominique, a 50 years French tourist fascinated by the history and culture of the countries he visits. Dominique comes to Marrakech for the first time and uses a Mobile GIS (MGIS) as an assistant.

3.1 Limitation related to localization technologies

Dominique wants to taste the local gastronomic specialties. So he interrogates his MGIS to get the list of the closest restaurants. After choosing one of the proposed restaurants, the system displays the route and the navigation assistant guides him to get there. The planned route passes through the narrow streets of the old Medina, thus GPS coverage is reduced or even unavailable sometimes. This makes Dominique abandon the navigation system and ask passers-by his route."

This use case illustrates one of the most important challenges of GIS technology, which is the precise determination of location. In fact, location positioning is more and more important in many application services. The maturity of application services and enormous governmental support has driven the rapid advancement of navigation industry [32]. The location positioning constraint is mainly due to the limitations of satellite positioning systems such as GPS, GLONASS, and Galileo who do not work inside buildings, tunnels, or in high-density urban areas (old villages with narrow streets or major business centres with high buildings). Other location technologies also have this problem of geographical coverage. Consequently, GIS technology needs new approaches to determine continuously precise positioning.

3.2 Limitations related to data access by proximity

Dominique arrives near the selected restaurant. Before entering, he wants to check the menu of the day, as well as the popularity of this restaurant among former clients feedback. To do so, he tries to access a remote web service. Unfortunately, the Internet connection is not available;

Dominique abandons his MGIS and looks to be informed in a different way".

This use case illustrates the problem of data access and filtering by proximity. In fact for a mobile user, geographic queries concerns usually geographic objects around him. However, to obtain information and services related to these objects, the user must consult remote servers. Besides the connection problems with these servers, the spatial query results can be far from the user interest. The context-aware systems offer a solution to this, by taking into account user's profile for example. But this approach knows several limitations such as the extraction and formulation of the user profile. Thus, GIS technology needs new understanding of the physical proximity; it should be extended to be also a contextual and semantic proximity. And this involves new forms of interaction of GIS applications with the geographic environment.

3.3 Limitations related to geographic data update

"After a good meal, Dominique wants to return to his hotel. He consults the MGIS to find a quick and shortest way. The itinerary proposed by the system passes through Mohamed VI Avenue. Unfortunately, the system does not cognize that barriers block this avenue due to the Film Festival of Marrakech. Dominique discovers this information when the barriers obstruct his way. Although the MGIS can propose another itinerary to the hotel, if he had the information earlier, he would have avoided him to make a long detour".
This use case reflects the challenge of geographic data update, which is an essential step in the GIS life cycle. This is an important challenge because of the permanent natural and artificial events that change the geography of the earth. These changes become more frequent due to population growth and the rapid development of urban areas. Accordingly, geographic data providers should update their databases and all associated applications continuously. However, the frequency whereby the earth changes exceeds the frequency of geographic data update. Hence, GIS technology needs new geographic data update approaches.
Previous use cases illustrate some of our research's motivations. Before presenting the response of Smart Geographic Object to these limitations, the following section details a conceptual architecture of Smart Geographic Object functioning.

## 4. Conceptual architecture of smart geographic object

The following architecture (Figure 4) shows in detail the components of the smart geographic object and its interactions with the environment, which are classified to proximity and remote interactions.

Smart Geographic object includes the geographic object and the hardware and software components that allow its instrumentation to communicate with its environment. Sensors attached to the geographic object are at the head of these components (1). They cooperate to determine the state of this object (location, temperature, motion, etc.). They are the first brick in the bridge between the physical and the digital worlds. The sensors are connected with the description of the geographic object (2) through a software layers, and they feed this description continuously by the data reflecting the geographic object state.

As mentioned before the description of the geographic object (3) contains geometric and thematic data plus the related services. To exchange this description, the concept of smart object is established. And it will be implemented through a smart device attached to the geographic object (5), this smart device plays the role of transceiver, and supports the nearby wireless communication technologies as well as conventional communication technologies.

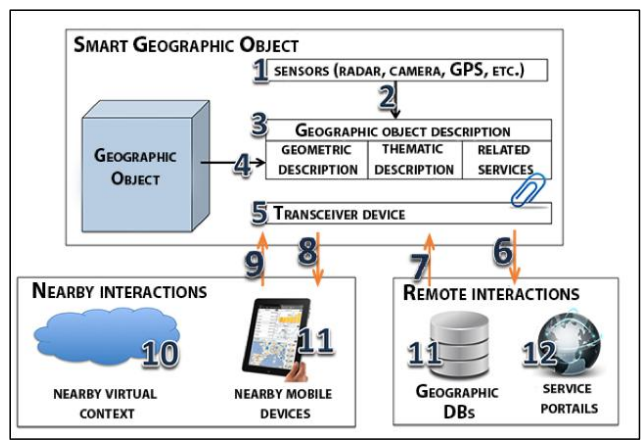

Fig 4 Conceptual architecture and basic components of the smart geographic object

4.1 Nearby interactions

In the Internet of things, the functionality offered by smart objects (e.g., the provisioning of online sensor data) is often referred to as real-world services. They are called so because they are provided by embedded systems related directly to the physical world [40]. In our work, nearby interactions reflect the real-world services offered by the smart geographic object to its closeness-surrounding environment. These services can be offered via SOAP-based web services or RESTful architectural design [33].

Nearby interactions represents the exchanges between SGO and surrounding mobile devices (11) such as smart phones or other smart geographic objects. In fact a smart geographic object can be requested by a set of mobile devices that appear and disappear in the diameter of interaction. This reminds an important challenge in ubiquitous computing which is service discovery. Hence a discovery policy can be established to enable smart geographic objects introduce themselves and publish their services. In this context, many service discovery protocols have been proposed such as SLP, HAVi, UPnP and Jini.

As an example to nearby interactions, a navigation assistant can access and retrieve SGO description like localization to be used in the navigation. Consequently, the smart geographic object can be an alternative source of information if the access to remote resources is not available.

In turn mobile devices can join the interaction diameter and discover the services offered by the smart geographic objects. Mobile devices - their side- can enrich smart geographic object description by new information. This part will be developed in our future work. We envisage constituting through the information provided by nearby devices a "Contextual Memory". The role of this Contextual Memory is to store former interactions history with mobile devices. For example the content of this can be passer-by feedback about a restaurant.

4.2 Remote interactions

Remote Interactions reflect exchanges between smart geographic object and remote servers. In one direction (9), remote servers send information and services that update corresponding entries of the smart geographic object description. Remote interactions enable the geographical communicating object to be aware of the remote changes concerning it. For example, the territorial division is information determined by the official authorities. Geographic servers can update this information in the smart object description trough remote interactions. Another way, smart geographic objects provide information to the remote servers. This information is provided permanently and continuously. Hence an instantaneous update of the remote servers is performed.

This conceptual architecture has outlined our approach and a minimum of components required to implement a smart geographic object. In the next part we present the interest to do this implementation by explaining the contribution of the smart geographic objects.

## 5. Contribution of the Smart Geographic Object concept

In the motivation scenario, we have identified three use cases where GIS technology shows its limits. We return to this scenario to illustrate the contribution of smart geographic object. It's about new forms of interaction with the geographical environment that meet the identified limitations:

5.1 Response to localization technologies limitations

As a response to this limitation, the concept of smart geographic object is not an absolute alternative, but an additional solution to refine the positioning. Actually, a smart geographic object can help improve localization, or offer an independent solution if satellite positioning solution are unavailability. Basically, Smart geographic objects share information about their location. Systems and applications can use this information to purify the given satellite positioning. They can also use this information to have an initial location positioning. In the latter case, a mechanism similar to the one used by satellites may be used, a set of smart geographic object cooper to triangulate the position. This was subject of several research works in indoor environments [29].

On the other hand, localization through smart geographic object can be more reliable. This reliability is due to the relative safety of proximity communication technologies like Bluetooth and ZigBee. Furthermore, there is a diversity of smart geographic objects, which ensures continuity and constant availability of the location information.

Our motivation scenario can be extended as follows: "Dominique's MGIS recognizes the unavailability of GPS coverage and start listening to the smart geographic object nearby. The Atlas Hotel provides its location to the MGIS; the navigation system resumes the assistance and consults smart geographic objects nearby regularly to update the location information. "

5.2 Response to data access by proximity limitations

Smart geographic object can allow a new approach of data storage. Instead of being exclusively stored in central servers, data can also be stored in the objects of surrounding environment. This ambient data storage upsurge a new approach of data filter by proximity. In fact, when a mobile user stops near geographic object, this means that this object is interesting for him. So this object can provide locally and implicitly relevant information to this user.

Other way, the potential profile of the user can be determined from the description of the smart geographic object. For example if a navigation system detect in the same session several interactions with "Historical SGO", it will recognize that the user profile is "History and Culture". In addition to that, smart geographic object can even inform the status or the needs of a mobile user. For example a tourist, who stops 3 times in front of nearby restaurants, is probably hungry and wants to take a meal.

Accordingly, Smart Geographic Object could sustenance context-aware systems to determine the user profile and state. Many ontology-based user profile modeling has been proposed like COBRA-ONT [34], OntobUM [35], Unified User Context Model (UUCM) [36] and UPOS [37]. As example of ntologyontology-based user profile modelingmodelling, figure 5 presents OntologieOntology Top Level Classes defined by Skillen et Al [38].

| Class Name | Class Description | Example Values |
|---|---|---|
| User | Type of user involved. | "Physically-disabled", "Normal" |
| User_Profile | Every user has one user profile. | |
| Activity | User activities, social or work related. | "Stamp collecting", "Writing" |
| Activity_Type | The type of activity the user is involved in. | "Indoor" "Cooking", "Outdoor" |
| Capability_Level | The level the user can cope with things. | "Severe", "Low", "Moderate" |
| Capability_Type | The type of capability that the user has. | "Cognitive", "Emotional" |
| Interest | A user hobby or work-related interest. | "Swimming", "Reading" |
| Interest_Type | The type of interest that the user has. | "Computing", "Food", "Sports" |
| Interest_Level | The level of interest the user has. | "Low", "Medium", "High" |
| Preference_Domain | The area of preference/likes. | "Exercise", "Food", "Travel" |
| Health_Condition | The health conditions associated. | "Dementia", "Diabetes" |
| Health_Level | The current health level/status of the user. | "Mild", "Normal", "Severe" |
| Location | Where the user in situated at one time. | "House", "Park", "Cinema" |
| Time | The time of day associated. | "Morning", "Evening" |
| Context | The environment that the user is | "Working", "Social", "Home" |

Fig 5 User Profile Ontology Top Level Classes

In collaboration with SGO the values of the following Top Level Classes can be determined:

- Activity_Type: we can determine if user is in outdoor or indoor, and depending on the current SGO description we can have an idea about the user current activity.
- Location: the user location can be determined through to position of the nearby SGO.
- Interest, Interest type, Preference_Domain: the SGO description can inform - even relatively - the interest of the user; this description can be combined with other resources to have a more precise user profiling.

The motivation scenario can be extended as follows: "Dominique desires to consult the menu of the day and the popularity of the restaurant. Unfortunately the Internet connection is not available. As a smart geographic object, the restaurant shares an e-menu linked directly to the restaurant management system. In the virtual nearby context, a set of contextual notes inform Dominique's MGIS about the restaurant popularity. The contextual notes contain each one a rank by a former visitor. Dominique is convinced by this restaurant, so he took his meal there."

5.3 Response to spatial data updates limitations

Remote interactions of the smart geographic object allow it to regularly exchange its description. Each change in the geographic properties will be detected by the sensors, reported in the description and forwarded to geographic databases. Similarly, a local update of the related services leads to renew information in the remote service portals. Thus Smart geographic objects enable a continuous, automated update, for geographic database as well as services portals. This also allows a relative validity of data; because information is given directly by the geographic environment.

The motivation scenario in this case, will be extended as follow: "Returning to the Hotel, Dominique consults the MGIS to find his path. The shortest route goes through Mohamed VI Avenue. Due to the Marrakech Film Festival, barriers block this avenue. As smart geographic object, these barriers send their positions to geographic servers. Geographic servers receive the position of barriers and declare Mohamed VI Avenue an impasse. So the MGIS takes this information into account and offers to Dominique a route through the Wintering Avenue which is parallel".

# 6. Related works

Belonging to different disciplines, the works we studied in the literature have the same aim; it consists on making the interactions with geographic environment more intelligent. In order to do so, each work proposes an application of the ubiquitous computing in the geographic area. But this application takes different forms; each form reflects an understanding of the ubiquitous computing vision. For example some works [23] [16] [39] limited the application of ubiquitous computing in the geographical area to mobility and pervasive access to GIS applications. In addition to the mobility, other works [18] [19] [20] have added context-awareness to customize the results of geographical questions.

The related works we present in this section can be placed in a third category. A category of works that introduce the concept of instrumented geographical environment. And propose new forms of interaction with this environment. At the end of this presentation, we discuss the similarities and differences of these works in comparison to our work.

6.1 Related work presentation

The first work [21] provides an interactive framework that integrates the RFID technology and GIS technology to acquire and manage real-time information on the movement of pilgrims. The information is captured and stored at the end to extract the average speed and travel time of the bus. This work implements an RFID system that consists of RFID tags embedded in vehicles carrying pilgrims one side; and RFID readers installed in specific locations of each course other side. The data collected by RFID readers are sent via Wifi to be analysed and archived in dedicated servers. These data will be presented in GIS applications such as interactive maps, and will be used to optimize future planning of road traffic.

The second research work is a part of UCPNAVI project [22]. This project investigates the ability of a smart environment to provide location-based information. It also studies how this information can complement or replace traditional localization techniques. To do so, the authors analyzed sensors and appropriate methods used for positioning in ambient environments. This analysis tested the relevance, reliability and accuracy of these sensors and methods. The investigation then focused on the evaluation of RFID positioning in combination with wireless networks and reckoning in indoor environments. In outdoor environments, RFID / GNSS / DR integration is designed to determine the position.

The third work [25] proposes an ambient system that allows effective inspection of road infrastructures. The proposed architecture integrates RFID technology and geographic information systems GIS. It is based on four elements, (1) the attachment of RFID tags in road infrastructures (2) mobile phones integrating RFID readers (3) Framework for the Integration of RFID applications and GIS that the authors called ITAG server (Integration of e -Tag Application and GIS server), (4) a database that store inspection data and photographs of the sites inspected. The types of road infrastructures defined are bridges, sidewalks, tunnels, and road furniture (traffic lights, road safety mirrors, etc.). The objective of this work is the inspection of the infrastructure to trigger reparation instructions. To do so, the RFID tags are assigned to each target infrastructure, and RFID readers attached to staff's mobile phones. At each inspection mission, RFID readers read the RFID tag identification number of the target road infrastructure. This number is used as a cooperation key with the GIS through the Framework ITAG. On the one hand the staff can get information about the infrastructure in question with reference to GIS via the identification number. On the other hand the inspection data and pictures are transferred to a geographic database via the Internet. The administrator in central office confirms the received data and initiates reparation instructions if needed.

The fourth and last work [26] proposes a system to supervise energy and environmental parameters of a building. The system is developed based on distributed sensors that communicate through ZigBee technology. The sensors allow the collection and monitoring of different types of measures. Two groups of sensors are proposed, the first group supervises energy through regular digital counters for energy. These counters calculate the consumption of electricity and gas. The second group supervises environmental parameters. This group of sensors includes a set of wireless sensors deployed in the rooms to measure temperature, humidity, noise, $CO_2$ concentration and air quality. The system was experienced on the Southeast University campus. A wireless sensors network using ZigBee technology has been implemented in some campus buildings. And a server was deployed to collect and store the measured parameters. Afterward, the sensors network capture and send continuously collected data to the server. Finally, the accumulated measurements are integrated into an application for centralized management and visualization. The system is integrated with a GIS application, which is a 3D map of the campus. This integration provides the supervisor an intelligible graphical interface that enables easy access to different rooms, floors and buildings of the campus.

6.2 Discussion

The first and second works propose the integration of RFID tags in already known locations. This Tags communicant with RFID readers connected to remote servers. Then an instant and precise location can be recovered. This approach constitutes a relative response to most of location technology limitations. The second work proposes a further combination with satellite tracking technology to determine the position more correctly in outdoor spaces. In our work we propose a rich description of the geographic object. Beyond the location, this description inform about the thematic and the related services to a geographic object. So in addition to problems related to the location, we offer an analysis and solutions to other problems related to interactions with the geographical environment such as updating and filtering data.

In the third work a GIS / RIFD integration is proposed. This integration allows extracting information about the spatial objects from the GIS by mobile users on the train. However, a break in the connection with the remote GIS causes a cessation on the functioning. This demonstrates the advantage of the ambient storage of information we propose; an ambient storage that enables data allocation and filtering by proximity. On the other hand, our automatic data update approach can provide an optimization to this work; instead of receiving inspection data exclusively from staff's smartphones, the road infrastructure can be equipped with sensor networks and each change in this infrastructure state trigger a repair notification to avoid random inspections and to perform well-localized instructions.

The fourth work introduced a system that collects data from the geographical environment through a sensors network. This joins our approach on the richness of instrumentation of the geographical environment as well as the richness of information to extract. This work also joined our approach on the instant data update; because the data collected by the sensors are sent automatically to the servers. However, the proposed system is limited to desktop data visualization and ignores the aspect of mobility and ubiquity of information that are the core concepts of the ubiquitous computing. On the other hand, the system restricts the integration with GIS to data visualization. GIS is actually used to access visually to locations through a 3D digital map.

From the analysis and evaluation of previous related work, we believe that our work involves:
- A rich and open theoretical framework to implement ambient computing in the geographic domain. This theoretical framework is based on the concept of the spatial object and the concept of communicating object and their fusion.
- A technologic independent and open conceptualization.
- A support for communication and updating various geometric and thematic information and services related to a spatial object.

## 7. Prospects and future work

We presented in this work, the concept of Smart Geographic Object. This concept came through a theoretical study to position GIS technology in ubiquitous computing era. We also presented the added value of SGO and illustrated how this concept can responds to some crucial GIS limitations. However, the current work rise a set of challenges that will be explored in the future works. First, the descriptions of the geographic objects constitute a real issue due to the complexity of the geographic environment. Since we are working on geographic object in urban area, we are currently investigating some geographic data model for the storage and the exchange of virtual 3D city models. The majority of these models have an XML-based format, which will make the extension of the meta-models possible to describe the geographic object our way. Second, our proposition is build on making the geographic environment a part of the GIS technology functioning. This means that the geographic environment will move from a state to a dynamic state. And will offer various functionalities. In ubiquitous environments the functionalities are more and more modeled and published as services. This leads us to think about a service oriented architecture that enables the discovery and the composition of the services offered by the Smart Geographic Object one side; and to manage the heterogeneity of these services on the other side. In this context several ubiquitous middleware are proposed, especially in Smart city projects.

Our current work open the way to many promising implementations - based on ubiquitous computing core concepts - especially in the context-aware mobile geographic systems. It also extends the functioning of GIS technology out side the system it self to make the environment an integral part of this functioning.